\documentclass[oldversion]{aa} 

\usepackage{graphicx}
\usepackage{txfonts}
\usepackage{natbib}
\bibpunct{(}{)}{;}{a}{}{,}

\newcommand{\ros}{{\em ROSAT}}

\newcommand{\chan}{{\em Chandra}}

\newcommand{\xmm}{{\em XMM-Newton}}

\newcommand{\eROS}{{\em eROSITA}}
\newcommand{\nh}{N_{\rm H}}

\def \msev{M7}
\def \magoneeig{RX J1856.5-3754}
\def \magzersev{RX J0720.4-3125}

\def \magtwoone{RX J2143.0+0654}

\def \zersixfiv{2XMM J104608.7-594306}
\def \candzsf{XMM J1046}

\begin{document}

\title{The isolated neutron star candidate \zersixfiv}

\author{A. M. Pires\inst{1,2}
        \and C. Motch\inst{2}
        \and R. Turolla\inst{3,4}
        \and A. Treves\inst{5}
        \and S. B. Popov\inst{6}
        \thanks{Based on the public data archive of \xmm\, an ESA Science Mission with instruments and contributions directly funded by the ESA Member States and the USA (NASA) and a \chan\ Legacy programme. Optical observations were performed at the European Southern Observatory, Paranal, Chile, under programme ID 079.D-0633(A), and at the Southern Astrophysical Research Telescope, Cerro Pach\'on, Chile.}}

\offprints{A. M. Pires}

\institute{
           Instituto de Astronomia, Geof\'isica e Ci\^encias Atmosf\'ericas, Universidade de S\~ao Paulo, R. do Mat\~ao 1226, 05508-090 S\~ao Paulo, Brazil,
           \email{apires@astro.iag.usp.br}
           \and
           Observatoire Astronomique, UMR 7550 CNRS, 11 rue de l'Universit\'e, 67000 Strasbourg, France
           \and
           Universit\'a di Padova, Dipartimento di Fisica, via Marzolo 8, 35131 Padova, Italy
           \and 
           Mullard Space Science Laboratory, University College London, Holmbury St. Mary, Dorking, Surrey, RH5 6NT, UK
           \and 
           Universit\'a dell'Insubria,  Dipartimento di Fisica e Matematica, Via Valleggio 11, 22100 Como, Italy
           \and
           Sternberg Astronomical Institute, Universitetskii pr. 13, 119991 Moscow, Russia
}

\date{Received ...; accepted ...}


\titlerunning{The neutron star candidate \zersixfiv}
\authorrunning{A. M. Pires et al.}

\abstract
{
Over the last decade, X-ray observations unveiled the existence of several classes of isolated neutron stars (INSs) which are radio-quiet or exhibit radio emission with properties much at variance with those of ordinary radio pulsars. 
The identification of new sources is crucial in order to understand the relations among the different classes and to compare observational constraints with theoretical expectations.
A recent analysis of the 2XMMp catalogue provided less than 30 new thermally emitting INS candidates. Among these, the source \zersixfiv\ appears particularly interesting because of the softness of its X-ray spectrum, $kT=117\pm14$\,eV and $\nh=(3.5\pm1.1)\times10^{21}$\,cm$^{-2}$ (3\,$\sigma$), and of the present upper limits in the optical, $m_{\rm B}\ga26$, $m_{\rm V}\ga25.5$ and $m_{\rm R}\ga25$ (98.76\% confidence level), which imply a logarithmic X-ray-to-optical flux ratio $\log(F_{\rm X}/F_{\rm V})\ga3.1$, corrected for absorption. We present the X-ray and optical properties of \zersixfiv\ and discuss its nature in the light of two possible scenarios invoked to explain the X-ray thermal emission from INSs: the release of residual heat in a cooling neutron star, as in the seven radio-quiet \ros-discovered INSs, and accretion from the interstellar medium.
We find that the present observational picture of \zersixfiv\ is consistent with a distant cooling INS with properties in agreement with the most up-to-date expectations of population synthesis models: it is fainter, hotter and more absorbed than the seven \ros\ sources and possibly located in the Carina Nebula, a region likely to harbour unidentified cooling neutron stars. The accretion scenario, although not entirely ruled out by observations, would require a very slow ($\sim$\,10\,km\,s$^{-1}$) INS accreting at the Bondi-Hoyle rate.
}

\maketitle

\section{Introduction}

A major outcome of the \ros\ mission has been the discovery of a group of radio-quiet isolated neutron stars (INSs) with properties clearly at variance from the bulk of the population of rotation-powered pulsars. At present, the group comprises seven objects (whence the nickname ``Magnificent Seven'', or \msev) sharing similar properties, which include purely thermal X-ray spectra with very soft temperatures ($kT\sim40$--100\,eV) and low absorption column densities ($\nh\sim$ few $10^{20}$\,cm$^{-2}$), long spin periods ($P\sim3$--10\,s), intense magnetic fields ($B\sim10^{13}$--10$^{14}$\,G), very faint optical counterparts ($m_{\rm B}\ga25$ when detected, implying X-ray-to-optical flux ratios $F_{\rm X}/F_{\rm{opt}}\ga10^4$) and no associations with supernova remnants or stellar companions \citep[see e.g.][for a recent review]{hab07}.

For some time, following the pioneering work by \cite{ors70} and \cite{shv71}, it was thought that these sources might belong to the long sought population of old INSs powered by accretion of the interstellar medium (ISM), which was expected to be detected by \ros\ \citep[see][for a review]{tre00}. However, the measurement of large proper motions \cite[e.g.][and references therein]{mot07,mot08} strongly supports the idea that the \msev\ represent a local population ($d\la500$\,pc, \citealt{pos07,ker07a}) of middle-aged neutron stars ($\sim$\,10$^5$--10$^6$\,yr), which give off thermal radiation as they cool down. The similar spin periods and their position on the $P-\dot{P}$ diagram raised the possibility that they could be related to other populations of INSs -- like magnetars, high magnetic field radio pulsars (HBPSRs) and rotating radio transients (RRATs; \citealt{lau06}) -- which is an intriguing and still debated issue \citep[e.g.][]{pop06,kap08,kea08}.

\begin{table*}
\begin{center}
\begin{tabular}{l c c c c c c c l c} 
\hline\hline
 OBSID & Observation Date & Detector & Filter & $t_{exp}$ & $\theta$ & $t_{eff}$ & Counts & Rate & Near Gap?\\
       &                  &          &        & (ks)      & (arcmin) & (ks)      &        & (10$^{-2}$\,s$^{-1}$) & \\
\hline
 112580601 & 2000-07-26 & pn & thick  & 27.7 & 9.7  &  15.0 & 558  & $3.73\pm0.18$ & yes\\
           &            & M1 & thick  & 33.2 & 8.7  &  17.8 & 244  & $1.37\pm0.10$ & no \\   
           &            & M2 & thick  & 30.1 & 9.1  &  15.7 & 198  & $1.26\pm0.10$ & yes\\
 112580701 & 2000-07-27 & pn & thick  & 8.0  & 9.7  &   4.3 & 161  & $3.7\pm0.3  $ & yes\\
           &            & M1 & thick  & 10.9 & 8.7  &   5.9 &  88  & $1.50\pm0.19$ & no \\
           &            & M2 & thick  & 7.9  & 9.1  &   4.1 &  41  & $1.00\pm0.18$ & yes\\
 112560101 & 2001-06-25 & pn & thick  & 21.5 & 15.8 &   6.7 & 291  & $4.4\pm0.3  $ & yes\\
 112560201 & 2001-06-28 & pn & thick  & 20.5 & 15.9 &   6.3 & 200  & $3.2\pm0.3  $ & yes\\
 145740101 & 2003-01-25 & M1 & thick  & 6.9  & 8.3  &   4.6 &  78  & $1.68\pm0.22$ & yes\\
 145740201 & 2003-01-27 & M1 & thick  & 6.9  & 8.3  &   4.6 &  73  & $1.59\pm0.21$ & yes\\
           &            & M2 & thick  & 6.9  & 8.0  &   4.8 &  69  & $1.44\pm0.20$ & yes\\
 145740301 & 2003-01-27 & M1 & thick  & 6.8  & 8.3  &   4.5 &  59  & $1.30\pm0.20$ & yes\\
           &            & M2 & thick  & 6.8  & 8.0  &   4.7 &  68  & $1.44\pm0.19$ & yes\\
 145740501 & 2003-01-29 & M2 & thick  & 6.9  & 7.9  &   4.8 &  67  & $1.41\pm0.19$ & yes\\
 160160101 & 2003-06-08 & M1 & thick  & 16.5 & 8.5  &   8.8 & 127  & $1.44\pm0.15$ & yes\\
           &            & M2 & thick  & 15.5 & 8.2  &   8.6 & 114  & $1.33\pm0.15$ & no \\
 160160901 & 2003-06-13 & M1 & thick  & 31.1 & 8.5  &  16.6 & 207  & $1.25\pm0.10$ & no \\
           &            & M2 & thick  & 31.1 & 8.3  &  17.1 & 217  & $1.27\pm0.10$ & no \\
 145780101 & 2003-07-22 & M1 & thick  & 8.4  & 9.4  &   4.6 &  76  & $1.66\pm0.21$ & no \\
 160560101 & 2003-08-02 & M2 & medium & 11.8 & 9.1  &   6.3 & 121  & $1.93\pm0.19$ & no \\
 160560201 & 2003-08-09 & M1 & thick  & 12.2 & 8.6  &   6.8 & 105  & $1.54\pm0.17$ & no \\
           &            & M2 & medium & 12.1 & 9.1  &   6.4 &  96  & $1.50\pm0.18$ & no \\
 160560301 & 2003-08-18 & M1 & thick  & 18.5 & 8.6  &  10.8 & 155  & $1.42\pm0.13$ & no \\
           &            & M2 & medium & 18.5 & 9.4  &  10.0 & 160  & $1.60\pm0.14$ & no \\
 206010101 & 2004-12-07 & pn & medium & 19.3 & 17.3 &   5.8 & 366  & $6.3\pm0.4  $ & no \\
 311990101 & 2006-01-31 & pn & thick  & 24.3 & 7.7  &  15.9 & 837  & $5.27\pm0.26$ & no \\	
           &            & M2 & thick  & 65.3 & 8.2  &  44.2 & 731  & $1.65\pm0.15$ & no \\
 9488      & 2008-09-05 & ACIS-I & OBF & 60.0 & 6.7 & 57.0  & 567  & $0.95\pm0.04$ & no \\
\hline
\end{tabular}
\caption{Description of the set of \xmm\ and \chan\ observations that serendipitously detected source \candzsf. The exposure times ($t_{exp}$), filtered for background flares, and the effective exposures ($t_{eff}$) on \candzsf, accounting for vignetting, are reported; $\theta$ is the source off-axis angle. Net source photons and the on-axis count rates are in the 0.15--3\,keV (\xmm) and 0.5--3\,keV (\chan) energy bands.\label{tab_Xdata}}
\end{center}
\end{table*}
Considering that within 1\,kpc the \msev\ appear in comparable numbers as young ($\la$ few Myr) radio and $\gamma$-ray pulsars \citep{pop03}, they may represent the only identified members of a large, yet undetected, elusive population of radio-quiet and thermally emitting INSs. The discovery of new candidates is then mandatory in order to make any progress towards an understanding of their properties as a population and of their relations with other classes of Galactic INSs.

Despite the relatively small field of view and low sky coverage of the \xmm\ Observatory\footnote{In a near future, \eROS\ will become an excellent tool to search for INSs (\texttt{http://www.mpe.mpg.de/projects.html\#erosita}).}, its large effective area and good positional accuracy at soft X-ray energies make it ideal to look for faint INS candidates. Recently, \cite{pir08} reported on the preliminary results of a programme aimed at identifying new thermally emitting INSs in the 2XMMp catalogue\footnote{\texttt{http://xmmssc-www.star.le.ac.uk/Catalogue/xcat\_\\public\_2XMMp.html}}. This version of the catalogue, released on July 2006, contains more than 120 thousand sources, covering a total sky area of $\sim$\,285 deg$^2$.
Out of the considered $\sim$\,7.2$\times10^4$ EPIC pn sources with count rates above 0.01\,s$^{-1}$, less than 30 good candidates met all the selection criteria -- i.e. well detected point-like sources showing soft thermal spectrum ($kT\le200$\,eV and $\nh=10^{19}-10^{22}$\,cm$^{-2}$), with no optical candidates and no identifications in over 170 astronomical catalogues. The brightest and most promising INS candidate is source \zersixfiv\ (hereafter \candzsf). In this work we discuss the properties of \candzsf\ in connection with the two main options concerning its nature, a cooling or an accreting INS. The description of the selection procedure and the results of follow-up optical observations on a sample of the X-ray brightest INS candidates will be presented elsewhere (Pires et al., in preparation).

\section{\candzsf\ in X-rays and the optical}

\subsection{Observations and data reduction}
Thanks to the fact that source \candzsf\ is at an angular distance of only $\sim$\,8.5\,arcmin from the well studied binary system Eta Carinae, it was serendipitously observed by the EPIC pn and MOS detectors on board \xmm\ on many different occasions. 
As part of an observational campaign to study the star-forming region of the Carina Nebula, \candzsf\ was also serendipitously observed once in a recent \chan\ observation (September 2008). Although this X-ray mission has observed Eta Carinae many times, \candzsf\ was not detected before given the smaller field of view of the ACIS instruments relative to the EPIC cameras.

\begin{table*}
\begin{center}
\begin{tabular}{l c | c c c c c c | c c c c} 
\hline\hline
\multicolumn{2}{c|}{} & \multicolumn{6}{c|}{Absorbed blackbody fit} & \multicolumn{4}{c}{Constant $\nh$}\\
\cline{3-8}\cline{9-12}
OBSID & Detector & $\nh$ & $kT$ & $f_{\rm X,14}$ & $C$ & d.o.f. & Goodness & $kT$ &$C$ & d.o.f. & Goodness\\
 & & (10$^{21}$\,cm$^{-2}$) & (eV) & (erg\,s$^{-1}$\,cm$^{-2}$) & & & (\%)& (eV) & & & (\%)\\
\hline
 112580601 & M1       &  $3.7^{+1.4}_{-1.1}$ & $132^{+16}_{-17}$ & $10.4(8)$	        & 19 & 21 & 43.7 & $134(5)$        &  19 & 20 & 40.6\\
 112580701 & M1       &  $3.8^{+2.2}_{-2.1}$ & $124^{+35}_{-23}$ & $11.6^{+1.5}_{-1.4}$ & 15 & 21 & 11.2 & $128_{-8}^{+9}$ &  15 & 20 & 12.6\\
 160160101 & M2       &  $5.6^{+3  }_{-2.3}$ & $103^{+26}_{-21}$ & $ 9.9^{+1.1}_{-1.0}$ & 19 & 13 & 81.6 & $125_{-6}^{+7}$ &  20 & 12 & 87.2\\
 160160901 & M1 / M2  &  $3.1^{+0.8}_{-1.0}$ & $132^{+17}_{-12}$ & $10.3(6)$	        & 25 & 32 & 13.1 & $126(4)$        &  25 & 31 & 16.0\\
 145780101 & M1       &  $2.7^{+2.2}_{-2.0}$ & $124^{+43}_{-26}$ & $12.9^{+1.7}_{-1.6}$ & 15 & 24 & 3.6  & $114_{-7}^{+8}$ &  16 & 23 & 4.0\\
 160560101 & M2       &  $5.9^{+2.6}_{-4}$   & $96^{+45}_{-23}$  & $11.5(1.2)$	        & 34 & 37 & 25.0 & $121_{-6}^{+7}$ &  35 & 36 & 25.7\\
 160560201 & M1 / M2  &  $1.4^{+1.1}_{-0.9}$ & $160^{+28}_{-23}$ & $11.6(9)$	        & 53 & 42 & 79.5 & $122(5)$        &  56 & 41 & 88.8\\
 160560301 & M1 / M2  &  $6.0^{+2.3}_{-1.1}$ & $ 92^{+ 9}_{-14}$ & $10.5^{+0.7}_{-0.8}$ & 33 & 30 & 44.9 & $115_{-3}^{+4}$ &  37 & 32 & 65.9\\
 206010101 & pn       &  $4.0^{+1.2}_{-1.0}$ & $106^{+14}_{-12}$ & $ 9.0(6)$	        & 35 & 26 & 86.7 & $112(4)$        &  35 & 25 & 89.1\\
 311990101 & pn / M2  &  $3.9^{+0.7}_{-0.5}$ & $122(8)$          & $10.1(3)$	        & 62 & 65 & 38.1 & $127(2)$        &  63 & 64 & 43.0\\
 9488      & ACIS-I   &  $6.3^{+0.7}_{-1.2}$ & $103^{+9}_{-8}$   & $ 7.8(3)$            & 13 & 20 & 12.9 & $125_{-3}^{+4}$ &  18 & 18 & 37.7\\
\hline			 
\end{tabular}
\caption{X-ray spectral analysis of INS candidate \candzsf. Only the set of 10 best \xmm\ observations (see text) and the \chan\ one are shown. The best fit parameters (derived using the maximum likelihood $C$ statistic) listed are for an absorbed blackbody model for which all parameters were allowed to vary freely (middle column). The right column shows the blackbody temperatures obtained when the column density is held constant to the mean value, $\nh=3.5\times10^{21}$\,cm$^{-2}$. Errors are 1\,$\sigma$. The observed flux $f_{\mathrm X,14}$ refers to ranges 0.15--3\,keV and 0.5--3\,keV for \xmm\ and \chan\ data, respectively, in units of $10^{-14}$\,erg\,s$^{-1}$\,cm$^{-2}$. The ``goodness-of-fit'' is derived from a number of 1000 Monte Carlo simulated spectra.\label{tab_Xfit}}
\end{center}
\end{table*}

We analysed 16 \xmm\ archival observations, spanning from July 2000 to February 2006, in which \candzsf\ was visible. The event files were processed applying standard procedures with SAS 7.1.2\footnote{\texttt{http://xmm.esac.esa.int/sas}}. The MOS and pn observations were reduced using \textsf{emchain} and \textsf{epchain}, respectively, applying default corrections. The event lists were filtered for intervals of high background activity as well as to retain the pre-defined patterns corresponding to single, double, triple and quadruple pixel events, for the MOS observations, and single and double pixel events, for the pn observations, as these have the best energy calibration. Source and background events were extracted using circular regions of radii $\sim$\,25\,arcsec (centered on the position of the X-ray source) and $\sim$\,50\,arcsec, respectively. Background regions were defined on an area free of sources in the same CCD and roughly at the same distance from the readout node as the source region. We restricted our analysis to the 0.15--3\,keV energy range.
Whenever possible for a given observation, data from all EPIC cameras were analysed simultaneously to better constrain the spectral parameters.

The \chan\ observation was analysed using CIAO 4.0.1 and CALDB 3.4.5\footnote{\texttt{http://cxc.harvard.edu/ciao4.0/index.html}}. We processed the event files with the task \textsf{acis\_process\_events}, also applying default corrections. Since the observation was taken in VFAINT mode, we cleaned the ACIS background while processing the event file. We checked for the presence of known processing offsets using the aspect calculator tool available on the CIAO web pages. Finally, we selected events corresponding to grades 0, 2, 3, 4 and 6 (in ASCA terminology) and applied the good time intervals, also filtering for periods of background flares. For the spectral analysis, we used the 0.5--3\,keV energy band due to the molecular contamination that is degrading the quantum efficiency of the ACIS front-illuminated chips, at energies below 0.5\,keV\footnote{\texttt{http://cxc.harvard.edu/proposer/POG}}. Source and background regions were extracted applying similar criteria as for \xmm\ data. In Table~\ref{tab_Xdata} we list, for every observation, the effective exposure times, off-axis angles, the number of extracted source photons for the spectral analysis and the source count rate, corrected for vignetting, in the given energy bands.

\subsection{Spectral analysis}
EPIC and \chan\ images do not reveal any particular background enhancement close to the X-ray source. However, the \chan\ data, which benefit from a lower instrumental background, do show some weak diffuse extended emission (a filamentary structure roughly consistent with some large-scale nebulosity seen in the optical\footnote{ESO MAMA-R digitized plate. This filament is not in the field of view of our ESO/SOAR images.}).
The brightest part is located 5 to 7\,arcmin south-east of the X-ray source. This filament is not detected on the \xmm\ images, although it is marginally in the field of view of the EPIC cameras. None of the background regions used in the spectral analysis overlaps with this diffuse X-ray emission.
Therefore, it is expected that the noise for the EPIC data is dominated by the (position independent) instrumental background. The noticeable vignetting and spreading of the PSF prevailing at the large off-axis angles where the source stands in many \xmm\ observations both contribute to the difficulty in measuring the spectrum of \candzsf\ in these cases.

Spectra were binned requiring a different minimum number of counts per energy bin, depending on the total number of source counts. Spectra extracted from the shortest exposures have at least 5 counts per bin.
Using XSPEC 12.4\footnote{\texttt{http://heasarc.gsfc.nasa.gov/docs/xanadu/xspec}}, we tested different models (blackbody, power law, bremsstrahlung, Raymond-Smith, \dots), allowing the fit parameters to vary freely. Due to the low number of counts, we applied the maximum likelihood $C$ statistic \citep{cas79} in order to derive the best fit parameters and their uncertainties. The quality of each fit (``goodness'') is estimated by means of Monte Carlo simulated spectra, drawn from the best fit model, and the distribution of corresponding $C$ fit statistics. If most ($>$\,50\%) of the simulated spectra has a smaller fit statistic than the current model, then it is unlikely that the observed data were drawn from the model. 

The spectrum of \candzsf\ is always best fitted by a single absorbed soft blackbody; other models (in particular, power law, optically thin thermal plasmas and magnetized neutron star atmospheres) invariably result in worse fits and there is no evidence for an additional (hard) component. 
In general, the X-ray emission of radio pulsars results from the sum of thermal and non-thermal components \citep[see e.g.][for a review]{kas04}. 
A power law component usually dominates the X-ray emission of young pulsars (age $\la$\,10$^5$\,yr) while old pulsars (age $\ga$ few Myr) exhibit a weak thermal component, probably originating from small heated polar caps, in addition to a dominating power law.
On the other hand, the X-ray emission of the middle-aged pulsars (age $\sim$ few $10^5$\,yr) known as ``Three Musketeers'' is clearly dominated by soft blackbody components superposed on a high energy tail with photon indexes 1.7--2.1 \citep[e.g.][]{luc05}. The addition of a power law component with similar photon indexes to the blackbody model does not improve the fit of \candzsf. Any power law component contributes at most 4\% (3\,$\sigma$ confidence level, 0.5--10 keV range) to the source luminosity. Therefore, a power law flux as low, relative to the thermal component, as that of the Three Musketeers (0.3--1.7\,\%) would not be detectable in \candzsf\ with the available data and can not be presently excluded. However, an X-ray spectrum dominated by a non-thermal component is clearly ruled out.

The \msev\ show rather stable spectral and timing properties.
In particular, the X-ray brightest source \magoneeig\ shows very constant flux and spectral properties over several years. Its X-ray spectrum is remarkably well reproduced by a blackbody with no significant deviations \citep[e.g.][]{bur03}. On the other hand, the second brightest source \magzersev\ is the only one among the \msev\ that has been shown to undergo long-term variations in its spectral parameters, at more or less constant flux. This behaviour is either possibly cyclic and related to the star precession \citep[][and references therein]{hab06,hoh08} or impulsive, being interpreted as a sudden change on the neutron star surface accompanied with a simultaneous torque, caused by an accretion or glitch episode \citep{ker07b}. 

Considering the 16 \xmm\ observations, single absorbed blackbodies have temperatures and column densities\footnote{The errors on $\nh$ reported in \citet[Table~2]{pir08} should be multiplied by a factor 10.} ranging from $76_{-11}^{+18}$ to $160_{-23}^{+28}$\,eV and from $\big(1.4_{-0.9}^{+1.1}\big)\times10^{21}$ to $\big(8.5_{-2.8}^{+2.6}\big)\times10^{21}$\,cm$^{-2}$; errors are 1$\sigma$. In some cases, the small number of counts prevents a well constrained fit. 
In order to check the significance of the variations of the spectral parameters of \candzsf, we performed a $\chi^2$ test on the 16 \xmm\ observations, assuming that they do not change in time and have values equal to their respective weighted means $kT = 113\pm11$\,eV and $\nh = (3.6\pm0.9)\times10^{21}$\,cm$^{-2}$ (hereafter, reported errors on the weighted means are 3\,$\sigma$, unless otherwise noted). We found that $kT$ and $\nh$ are consistent with being the same in all observations at an acceptable but low confidence level for the temperature ($\chi^2_{kT} \sim 26$ and $\chi^2_{\nh} \sim 19$ for 15 degrees of freedom, respectively; the probabilities of getting a larger $\chi^2$ are $\sim$\,4\% and $\sim$\,19\%).
However, we note that among the data sets with the largest values of $\chi^2$ there are some observations for which the source elongated point spread function (PSF) overlaps the CCD gaps of the EPIC instruments. The correction for the missing part of the energy-dependent PSF might not be well calibrated, in particular at large off-axis angles (\candzsf\ is located at $\sim$\,9\,arcmin, on average, but its off-axis angle can be as large as $\sim$\,15\,arcmin, see Table~\ref{tab_Xdata}).
This would require the inclusion of an additional systematic error which would have the effect of lowering the $\chi^2$ values. On the other hand, if we only consider the 10 observations for which the source is not located close to a gap\footnote{Observations flagged with ``no'' in Table~\ref{tab_Xdata}.}, we find that the temperature and column density are steady over the six-year time interval at rather high confidence levels: $kT = 117\pm14$\,eV (23\%) and $\nh = (3.5\pm1.1)\times10^{21}$\,cm$^{-2}$ (44\%), see Table~\ref{tab_Xfit}.

\begin{figure}[t]
\centering
\includegraphics[width=0.49\textwidth]{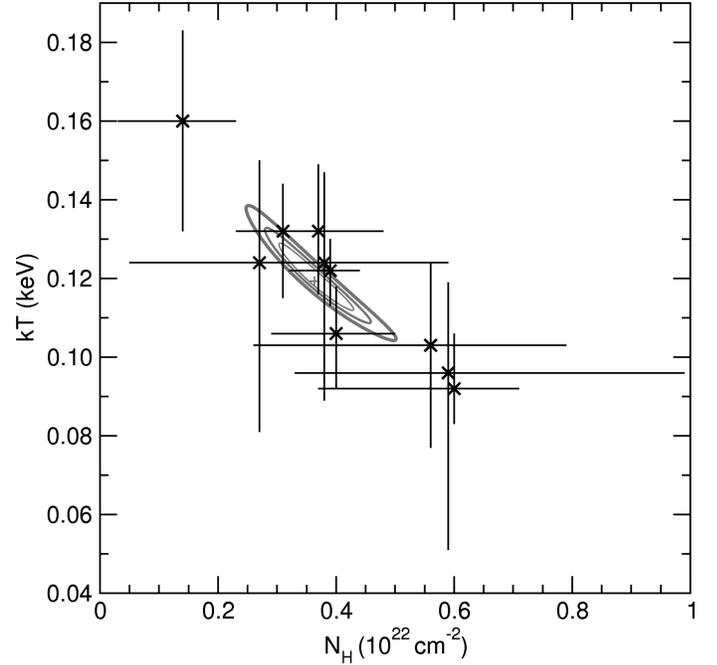}
\caption{Contour plot corresponding to the overall X-ray fit of source \candzsf, using the 10 best \xmm\ observations ($kT=122_{-8}^{+10}$\,eV and $\nh=3.6_{-0.7}^{+0.6}\times10^{21}$\,cm$^{-2}$; errors are 3\,$\sigma$). The $kT-\nh$ parameter space is shown for 1\,$\sigma$, 2\,$\sigma$ and 3\,$\sigma$ confidence levels. The individual best fits of Table~\ref{tab_Xfit} are shown as crosses with 1\,$\sigma$ error bars.\label{fig_contour}}
\end{figure}
\begin{figure*}
\centering
\includegraphics[width=0.49\textwidth]{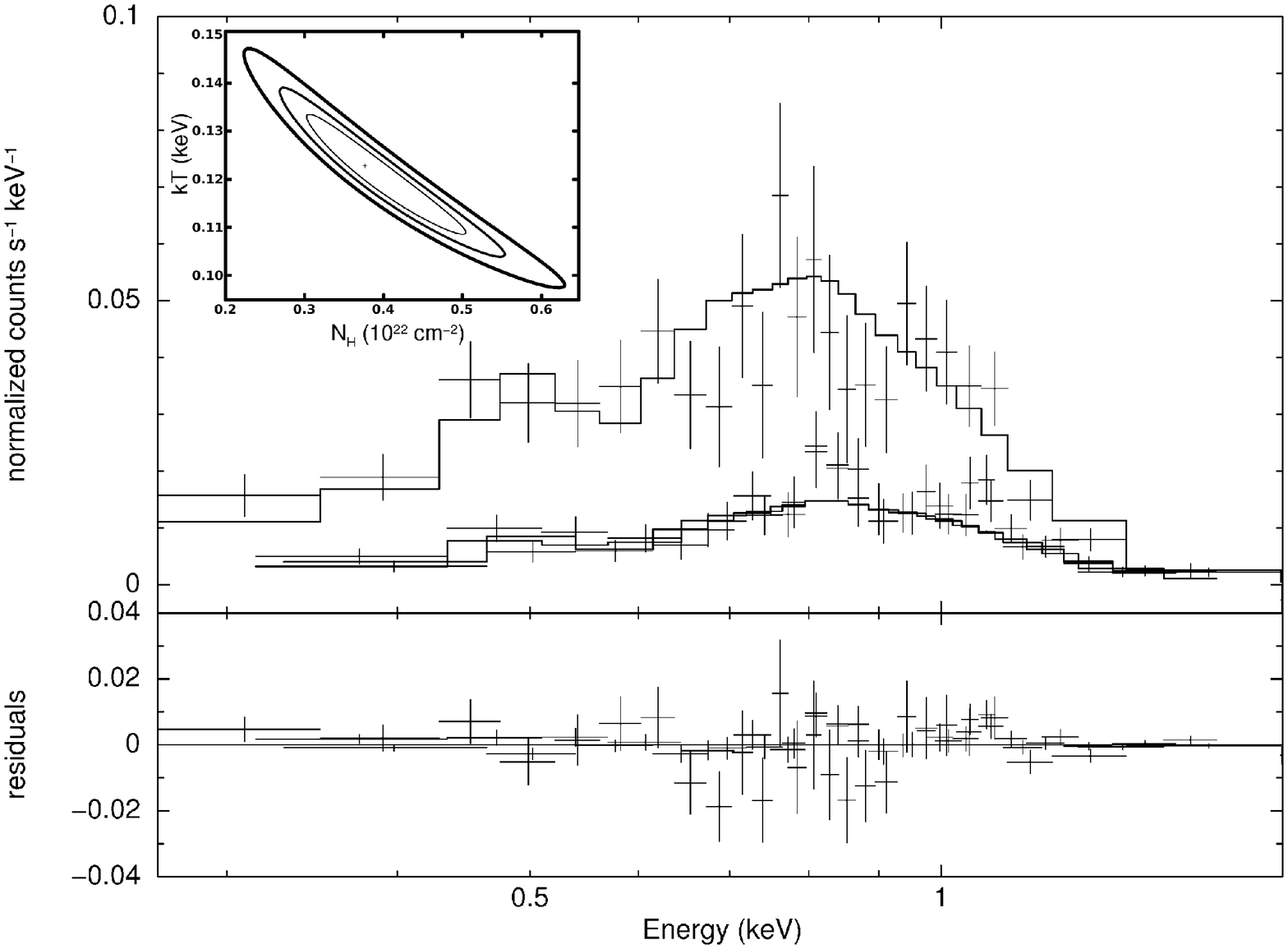}
\includegraphics[width=0.49\textwidth]{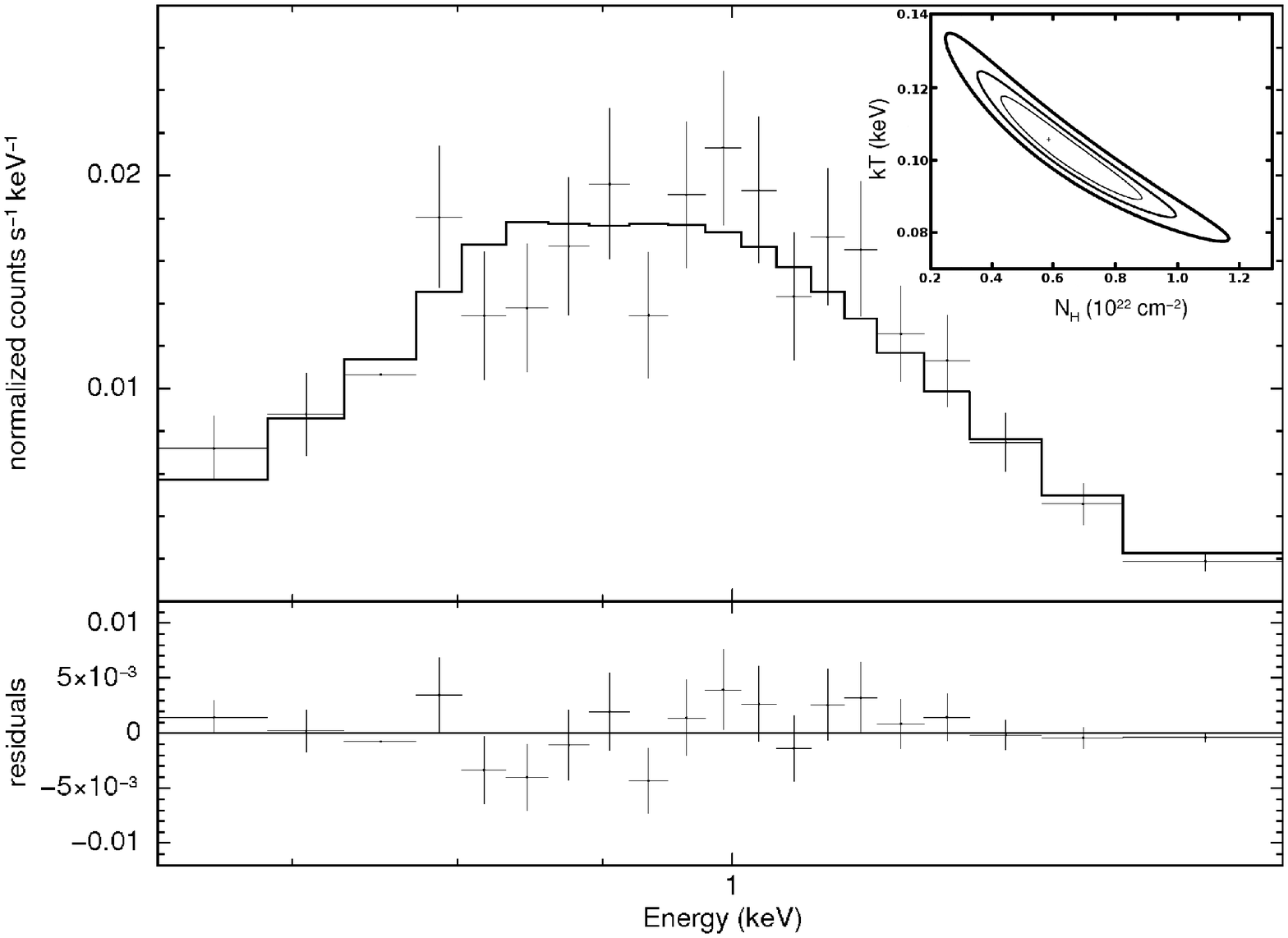}
\caption{X-ray spectra and best fits of source \candzsf\ as observed with \xmm\ (left, OBSID 311990101, pn and MOS2 cameras; the two lower data points consist of two subsequent MOS2 exposures) and \chan\ (right, ACIS-I). The contour plots in the insets show the $kT-\nh$ parameter space for 1\,$\sigma$, 2\,$\sigma$ and 3\,$\sigma$ confidence levels.\label{fig_spec}}
\end{figure*}

Alternatively, if the column density is held constant to the mean value $\nh = 3.5\times10^{21}$\,cm$^{-2}$, the computed range in $kT$ is narrower, 111--132\,eV, and the parameters are better constrained (1\,$\sigma$ errors correspond to $\sim$\,5\% of the best values against $\sim$\,20\% when $\nh$ is free to vary, see Table~\ref{tab_Xfit} for the individual fits). In this case, the variations in $kT$ are statistically significant (99.26\%). We note, however, that the sample of analysed observations is highly heterogeneous and thus subjected to systematic uncertainties. In particular, the different observing conditions -- distance to the optical axis, observing modes and cameras -- as well as calibration uncertainties at energies below 0.5\,keV especially for the MOS cameras\footnote{\texttt{http://xmm2.esac.esa.int/docs/documents/CAL-TN-\\0018.pdf}}, do not allow one to draw definite conclusions on the source intrinsic spectral variability. Hereafter, we adopt the weighted mean blackbody temperature and column density of the 10 best \xmm\ observations for the purpose of further discussion on the source properties. In Fig.~\ref{fig_contour} the individual best fits are plotted together with the $\nh\times kT$ contours obtained fitting all data with one single blackbody model.

A slightly more absorbed blackbody fit is found for the \chan\ observation (Table~\ref{tab_Xfit}). We note, however, that the response of the front-illuminated ACIS chips is not best suited to derive the spectral parameters of a source as soft as \candzsf. This, and the fact that only photons with energies above 0.5\,keV were considered, act together to cast some doubts on the column density derived in the blackbody fit. 
In Fig.~\ref{fig_spec} we show the spectral fits of the two data sets with the best signal-to-noise ratio among the analysed data. 

The weighted mean of the 0.15--3\,keV observed flux is $f_{\rm X} = (9.7\pm0.5)\times10^{-14}$\,erg\,s$^{-1}$\,cm$^{-2}$, considering the 16 \xmm\ observations; the computed range is $(7.7\pm0.7)\times10^{-14}$ to $\big(12.9_{-1.6}^{+1.7}\big)\times10^{-14}$\,erg\,s$^{-1}$\,cm$^{-2}$ (1\,$\sigma$). Although a constant flux is not statistically acceptable ($\chi^2 = 43$ for 15 degrees of freedom), a long term variability is excluded once the fluxes in the EPIC cameras are considered separately: the averaged flux in the pn camera is systematically fainter than that measured in the MOS cameras but both are consistent with a constant flux (at confidence levels of $\sim$\,19\% and $\sim$\,92\%, respectively). The same $\sim$\,13--20\,\% discrepancy between the two instruments is seen in the three observations where both cameras were simultaneously on. 
According to recent cross-calibration studies\footnote{S. Mateos talk in the \xmm\ Survey Science Center consortium meeting 2008, \texttt{http://xmm.esac.esa.int/external/xmm\_\\data\_analysis/ssc\_meeting/agenda.php}}, the MOS cameras register 7--9\,\% higher flux than pn below 4.5\,keV\footnote{The discrepancy between the cameras is even larger (of $\sim$\,12--13\%) at higher energies, according to the work just cited.} and this excess increases with off-axis angle. The larger discrepancy can thus be explained by the large off-axis angle of \candzsf\ in the set of analysed observations.
The probabilities that temperature and column density are constant also increase when the cameras are considered individually (a consequence of the larger errors), but there is no systematic relation as for the flux (i.e. the weighted mean values are roughly the same).

\begin{figure}[t]
\centering
\includegraphics[width=0.495\textwidth]{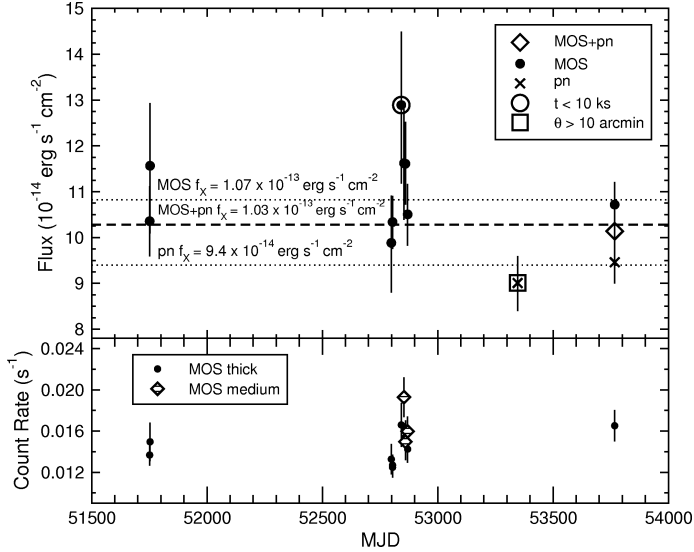}
\caption{\emph{Top:} Long term evolution of the observed flux (0.15--3\,keV) of the INS candidate \candzsf. Only the set of 10 \xmm\ observations not close to a gap is shown. Errors are 1\,$\sigma$ confidence level. The flux is stable in both pn and MOS detectors over the six-year time interval although there is a $\sim$\,13--20\,\% discrepancy between the measured fluxes in these instruments. Observations with short exposure times and for which the source is located at a large off-axis angle are highlighted. \emph{Bottom:} MOS on-axis count rates, discriminated by filter, for the same set of observations.\label{fig_flux}}
\end{figure}

Once again, accounting only for the set of 10 observations not close to a gap, the overall picture further argues against significant flux variations: the MOS and pn fluxes can be considered to be constant either when the cameras are analysed together or separately (Fig.~\ref{fig_flux}). The weighted means and respective confidence levels for the two cameras are $f_{\rm{X, pn}} = (9.4\pm1.1)\times10^{-14}$\,erg\,s$^{-1}$\,cm$^{-2}$ (44\%) and $f_{\rm{X, MOS}} = (1.07\pm0.08)\times10^{-13}$\,erg\,s$^{-1}$\,cm$^{-2}$ (78\%). Despite of the discrepancy between the two instruments, the fluxes are compatible within errors. We adopt here the mean value of the flux as measured by the two cameras, which is also consistent with the value inferred when the cameras are analysed together $f_{\rm X} = (1.03\pm0.06)\times10^{-13}$\,erg\,s$^{-1}$\,cm$^{-2}$. A direct comparison of the source count rate, corrected for vignetting, for the MOS cameras, is also shown in Fig.~\ref{fig_flux}.

\subsection{Timing analysis}
One of the most intriguing and discernible features of the \msev\ is that, when compared to radio pulsars, the neutron star spin periods are longer and distributed in a much narrower range. Six of the sources show sinusoidal X-ray pulsations with pulsed fractions between $\sim$\,1\% and 18\%. The detection of pulsations in the X-ray emission of source \candzsf\ would represent further evidence to unveil its nature. However, despite of the large number of archival observations that detected the source, only one (OBSID 311990101, pn) is really suitable to conduct timing analysis -- the others being either too short and having few source counts or, for the MOS observations, taken in full frame mode which has a too poor frame resolution (2.6\,s) for timing purposes. Similarly, the \chan\ data was carried out with a 3.4\,s time resolution imaging configuration.

With the aim to search for pulsations, we converted the photon arrival times of the pn observation from the local satellite to the solar system barycentric frame using the SAS task \textsf{barycen}. Source photons were extracted in a smaller elliptical region (relative to the extracted region for the spectral analysis) in order to avoid background events. For the same reason and, since \candzsf\ does not show counts above $\sim$\,1.5\,keV, we only considered the 0.15--2\,keV energy range. We next searched for pulsations using a $Z^2_n$ (Rayleigh) test \citep{buc83}. No pulsations are found to a non-constraining 30\% upper limit (3\,$\sigma$), in the 0.073--100\,s period range; the dimness of the source would require a longer exposure in order to significantly constrain the upper limit on pulsations.

\subsection{Optical follow-up}
Follow-up optical observations performed in the B and R bands at the ESO Very Large Telescope (VLT) in February 2007 revealed no counterpart within $\sim$\,4.3\,arcsec ($\ga$\,5\,$\sigma$) from the position of \candzsf, $\alpha=10$:$46$:$08.72$, $\delta=-59$:$43$:$06.4$, as derived running the SAS task \textsf{emldetect} on the best \xmm\ observation 311990101. The 90\% confidence level error circle on the position, $r_{90}=1.33$\,arcsec, is given by $2.15\sqrt{\sigma^2 + \sigma_{\rm{syst}}^2}$, where $\sigma$ is the nominal error as given by \textsf{emldetect} and $\sigma_{\rm{syst}}$ is the systematic error on the detection position as provided by the 2XMMp catalogue, based on reliable cross-correlation with the USNO B1.0 optical catalogue. Similarly, we used the CIAO task \textsf{wavdetect} to determine the position and the 90\% confidence level error circle using the \chan\ data, which is fully consistent with the \xmm\ results ($\alpha=10$:$46$:$08.7$, $\delta=-59$:$43$:$06.7$, $r_{90}=1.96$\,arcsec).

\begin{figure}[t]
\centering
\includegraphics[width=0.48\textwidth]{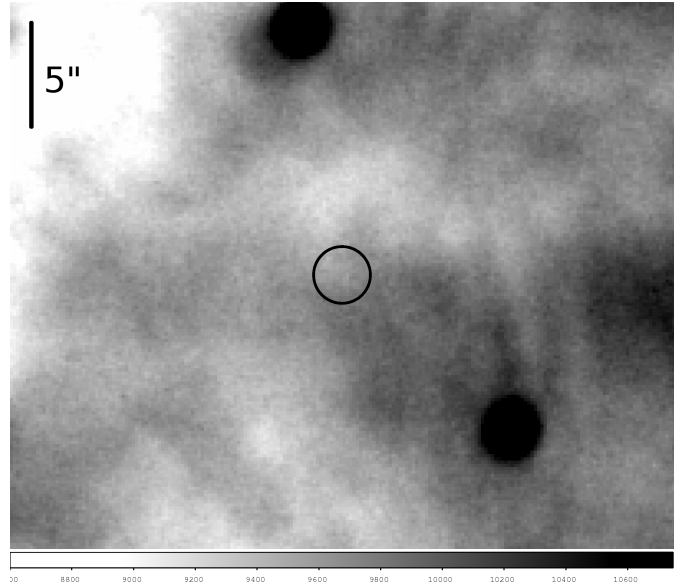}
\caption{Optical H$\alpha$ image of the field of \candzsf. An inverted colour map is used, i.e. brighter objects are darker. North and east point upwards and to the left, respectively. The image was smoothed using a Gaussian function (0.15\,arcsec). This SOAR observation is affected by fringing.\label{fig_opt}}
\end{figure}

The VLT observations consisted of short exposure (2$\times$150\,s in each filter) pre-imaging data which had the goal of selecting possible optical candidates for spectroscopy or deeper imaging\footnote{Details on the optical follow-up of this and other INS candidates are reported in a separate paper (Pires et al., in preparation).}. The optical images were astrometrically calibrated using non-saturated 2MASS and GSC-2 catalogued stars and the GAIA software\footnote{\texttt{http://star-www.dur.ac.uk/$\sim$pdraper/gaia/gaia.html}}. 
We obtained additional imaging on the source at the Southern Astrophysical Research (SOAR) 4.1\,m telescope in V and H$\alpha$ (6$\times$1120\,s and 6$\times$1575\,s) in February 2008, which also failed to detect any object. In order to compute upper limits for the optical counterpart of \candzsf, we defined the limiting magnitude of the data as the magnitude of the faintest simulated star still succesfully measured by means of PSF fitting, on the exact location of \candzsf. For all filters, the synthetic star was no longer detected or it was rejected (while trying to fit the PSF model) when the signal-to-noise ratio was worse than 2.5. At this confidence level (98.76\%), the present upper limits are $m_{\rm B}\ga26$, $m_{\rm V}\ga25.5$ and $m_{\rm R}\ga25$.
Although some structure in the H$\alpha$ nebular emission is present close to the position of the X-ray source, it is not clear if it is related to \candzsf\ since similar structures also appear in other parts of the nebula, see Fig.~\ref{fig_opt}.

After correcting for photoelectric absorption and interstellar extinction, the above limits imply $\log(F_{\rm X}/F_{\rm V})\ga3.1^{+0.3}_{-0.1}$. For an assumed $\nh=(3.5\pm1.1)\times10^{21}$\,cm$^{-2}$, $A_V\sim1.96(6)$ \citep{pre95} and the unabsorbed X-ray flux\footnote{In this work we adopted the notation $F_{\rm X}$ and $F_{\rm V}$ to denote the unabsorbed fluxes while $f_{\rm X}$ is used for the observed flux.} is $F_{\rm X}\sim\big(1.4^{+2.4}_{-0.8}\big)\times10^{-12}$\,erg\,s$^{-1}$\,cm$^{-2}$ in the 0.1--12\,keV energy band. The high value of the X-ray-to-optical flux ratio pratically rules out any other possibility than an INS. For instance, late-type (M, G, K) stars and active galactic nuclei (AGN) show logarithmic X-ray-to-optical flux ratios $\la$\,$-1$ and within $-1$ and $1$ \citep[e.g.][and references therein]{bar07}, respectively, while cataclysmic variable (CV) systems and BL Lac objects, which are among the most extreme classes of objects, have $\log(F_{\rm X}/F_{\rm V})\la2$ \citep[e.g.][]{schwope99}. Obscured T-Tauri stars would exhibit much more absorbed X-ray spectra. We also note that the high total Galactic extinction in the direction of \candzsf\ ($E(B-V) \sim 12$; \citealt{sch98}) rules out a background AGN. Moreover, the Parkes-MIT-NRAO Multibeam Survey of the southern hemisphere shows no radio source located at $\la$\,6\,arcmin from the position of \candzsf\ to a flux limit of $\sim$\,32\,mJy \citep{wri94}. The Parkes Multibeam Pulsar Survey has a sensitivity of about 0.14\,mJy for a canonical pulsar \citep{lyn08,man05} and also failed to detect the source -- the nearest entry in the ATNF Pulsar Catalogue is AXP 1E 1048.1-5937, at $\sim$\,32\,arcmin from \candzsf, and the young radio pulsar PSR J1052-5954, at $\sim$\,50\,arcmin.

This, together with its thermal, soft energy distribution, makes \candzsf\ a very promising INS candidate, with overall properties similar to the \msev. Deep radio searches and dedicated X-ray and optical observations should be carried out in order to confirm its nature and to determine to which subgroup of INSs it may belong.
For comparison, the identification of the \msev\ \magtwoone\ with an INS was first proposed on the basis of a $\sim$\,500 net counts \ros\ PSPC spectrum and of follow-up optical observations yielding $\log(F_{\rm X}/F_{\rm V})\ga3$ \citep{zam01}; the nature of the source was then confirmed by a dedicated \xmm\ observation \citep{zan05}.


\section{Discussion}

On the basis of the present data, the identification of \candzsf\ with an INS appears convincing. If proven correct, this would be the first example of a presumably radio-quiet and X-ray dim INS, located at a significantly greater distance than the \msev.

Most probably, all cooling INSs contained in the \ros\ Bright Source Catalogue have already been identified \citep{pop00a,rut03}, so that new sources must be looked for at lower fluxes. The search for ``blank field'' sources\footnote{Sources without an optical candidate in their X-ray error circles.} in the entire \ros\ All-Sky Survey and in HRI pointings \citep[e.g.][]{agu06,chi05,tre07} produced candidates with X-ray-to-optical flux ratios of $\sim$\,10 to 100. Further X-ray and optical investigations of one such blank field led to the discovery of Calvera, a likely INS (which exact nature is, however, still unclear) with a very large $F_{\rm X}/F_{\rm V}\ga8700$ \citep{rut08}. However, this source, like most of the other INS candidates discovered in the \ros\ data at faint fluxes, is at high Galactic latitude while it is expected that both accreting and cooling INSs at greater distances should be more abundant close to the Galactic plane.

Detailed population synthesis calculations \citep{pos08} show that, in general, cooling INSs at lower soft X-ray fluxes are also expected to be hotter than the \msev\ (an observational bias due to the higher photoelectric absorptions) and still at small angular distance from their birth star forming regions. Recently, \cite{mun08} used $\sim1000$ \xmm\ and \chan\ archival observations covering significant part of the sky close to the Galactic plane ($|b|\le5^\circ$) to present constraints on the number of magnetar candidates. The search was sensitive to sources with luminosities $L_{\rm X}\ge3\times10^{33}$\,erg\,s$^{-1}$ and pulsed fractions $p_{\rm f}\ge15$\,\%, at distances within a few kiloparsecs (see their Fig.~6). Interestingly, their results can also be used to constrain the number of cooling INSs, since no new pulsating neutron star candidate was found in spin period range $\sim$\,5 to 20\,s.
Rescaling luminosity down to the typical values inferred for the \msev\ ($L_{\rm X}\ge3\times10^{31}$\,erg\,s$^{-1}$) we obtain that the search presented by Muno et al. (2008) is sensitive to such objects up to a few hundred parsecs. Taking into account the expected distribution of cooling INSs on the sky \citep{pop05,pos08}, we obtain as a rough estimate that the limits by Muno et al. can be translated into $\la$\,1000 detectable cooling neutron stars up to a few hundred parsecs.
In comparison with other limit this does not provide new important constraints. Still, a similar search in slightly wider period range focused on the possibility of finding cooling INSs is welcomed.

The spatial location of \candzsf, close to the center of the Carina Nebula, and the derived value of the source column density suggest that it may be physically associated with this giant \ion{H}{ii} region. The Carina Nebula harbours a large number of massive stars and has ongoing active star formation (see e.g. \citealt[and references therein]{smi07}). The source column density is a factor of $\sim$\,10 higher than those typical of the \msev\ and is consistent with the one towards Eta Carinae ($\nh \sim 3\times10^{21}$\,cm$^{-2}$, \citealt{leu03}). Eta Carinae's distance, measured with high accuracy through the expansion parallax of its circumstellar nebula, is 2.3\,kpc (e.g. \citealt{smi06}). The distance to \candzsf\ is then likely to be comparable.

If the star surface is an isotropic blackbody emitter, then
\begin{displaymath}\label{eq_radius}
\frac{R_{\infty}}{d} \sim 3.05 \textrm{ } \Big(\frac{F_{\rm X}}{10^{-12}\rm\,erg\,s^{-1}\,cm^{-2}}\Big)^{1/2} \Big(\frac{kT}{100\rm\,eV}\Big)^{-2} \textrm{ km kpc$^{-1}$}
\end{displaymath}
which gives a radiation radius (as seen by an observer at infinity) of $R_{\infty} \sim 6.1$\,km for \candzsf, assuming that it is at the same distance as Eta Carinae. Although smaller than the canonical neutron star radius, such a value is in agreement with what is measured for the \msev: their redshifted radiation radii, as derived from X-ray blackbody fits and distance estimates, are in the range of $\sim$\,2 to 7\,km. However, it is well known that larger emission radii (and softer temperatures) are inferred when the surface radiation is described by more realistic models which take into account the overall spectra of these sources -- in particular, by invoking a geometrically thin hydrogen atmosphere on top of the condensed neutron star surface \citep[e.g.][]{mot03,zan04,hok07}. Moreover, the presence of small hot regions on the surface is understood in terms of the anisotropic heat transport that occurs in the crust of cooling neutron stars endowed with a strong toroidal magnetic field component \citep{per06,pag07}. Recent investigations of neutron star thermal evolution which account for these effects \citep{agu08} confirm that coolers can easily have polar caps with high temperatures and small radii. 

Alternatively, the rich star forming environment of the Carina Nebula brings in the intriguing possibility that \candzsf\ could be a much older neutron star accreting from the ISM, probably born outside the nebula and whose orbit is presently intersecting the \ion{H}{ii} region. We note that, in this case, position and velocity are not expected to be correlated.
The gas mass in the Carina Nebula is $\sim$\,10$^6$\,M$_\odot$ \citep{smi07} which, for a typical \ion{H}{ii} region size of $\sim$\,100\,pc, implies an average density of $\sim$\,10\,cm$^{-3}$. In fact, as reported by \cite{miz02}, two distinct electron density components are detected in a 30\,pc area centered on the Carina I and II \ion{H}{ii} regions: a high-density ($n_{\rm e} \sim$ 100--350\,cm$^{-3}$) component and an extended low-density ($n_{\rm e}\la100$\,cm$^{-3}$) component detectable over the entire mapped region. If \candzsf\ is moving inside the nebula, the corresponding increase on the column density is $\sim$\,1.5$\times10^{20}$\,cm$^{-2}$\,pc$^{-1}$, assuming a typical density of $\sim$\,50\,cm$^{-3}$. This means that the source could be up to $\sim$\,10\,pc inside the nebula and $\nh$ would still be compatible with the measured value and with the one derived for Eta Carinae.

At 2.3\,kpc, the luminosity of \candzsf\ is $L_{\rm X}\sim\big(9^{+15}_{-5}\big)\times10^{32}$\,erg\,s$^{-1}$. Although the estimated emission radius of \candzsf\ is comparable to those of the \msev, the higher blackbody temperature is responsible for the factor of nearly 10 higher X-ray luminosity.
If \candzsf\ is in the accretion phase, mass entrainment should then proceed at a rate $\dot{M} = L_{\rm X}/\eta c^2 \sim 5\times10^{12}$\,g\,s$^{-1}$, where $\eta\sim0.2$ is the efficiency. The Bondi-Hoyle accretion rate for a star moving through the ISM with particle density $n$ is $\dot{M}\sim10^{11} n (v_{10})^{-3}$ g\,s$^{-1}$, where $v_{10}$ is the velocity of the neutron star relative to the ISM in units of 10\,km\,s$^{-1}$. For $n\sim10-100$\,cm$^{-3}$, as appears likely inside the nebula, Bondi-Hoyle accretion can produce the required luminosity but the star should move very slowly through the gas, $v\sim10$\,km\,s$^{-1}$. Radio pulsars are known to have very high spatial velocity (typically $v\sim400$\,km\,s$^{-1}$; e.g. \citealt{hob05,fau06}), thus the chance to have an object so slow is very small. Morevover, several mechanisms are known to inhibit accretion onto a magnetized INS \citep{bla95,tor03,ikh07}. 

Interestingly, and in agreement with the results of simulations by \cite{pos08}, a higher temperature and greater distance respective to the \msev\ are also observed in the only RRAT (J1819-1458, \citealt{lau07}) detected up to now in X-rays. The \xmm\ spectrum of J1819-1458 is well fitted by a blackbody with $kT\sim140$\,eV and a broad absorption feature at $\sim$\,1\,keV -- similar to those of the \msev, which are usually interpreted as evidence for magnetic fields of $\sim$ few 10$^{13}-10^{14}$\,G. Its DM distance is $\sim$\,3.6\,kpc and the observed flux is $f_{\rm X}\sim1.5\times10^{-13}$\,erg\,s$^{-1}$\,cm$^{-2}$ (0.3--5\,keV). The \msev\ and this RRAT also share similar spin periods and period derivatives. On the other hand, steady or transient radio emission are not detected in any of the \msev\ to a rather sensitive limiting flux of $\sim$ few 10\,$\mu$Jy \citep{kon08}. In order to better understand the relations between these two classes of INSs, further, deeper pointings in X-rays and in radio are required to firmly assess the nature of this and other RRAT sources. 

\section{Summary and conclusions}

Overall, the present observational picture of \candzsf\ suggests that it has the right properties to be identified with a younger and more distant thermally emitting INS, possibly the eighth member of the \ros-discovered \msev. The analysis of its X-ray emission, although based on archival data obtained with non-optimal configurations, reveals an intrinsically soft energy distribution, possibly variable on long time scales. However, uncertainties on the calibration accuracy of the large off-axis angles at which the source was observed may account for the variations of the derived spectral parameters.
Its mean temperature, column density and 0.15--3\,keV observed flux are $kT=117\pm14$\,eV, $\nh=(3.5\pm1.1)\times10^{21}$\,cm$^{-2}$ and $f_{\rm X}=(1.03\pm0.06)\times10^{-13}$\,erg\,s$^{-1}$\,cm$^{-2}$, where we adopted the weighted means of the subsample of the 10 best \xmm\ observations.
The present optical limits imply a logarithmic X-ray-to-optical flux ratio greater than $\sim$\,3.1, already high enough to safely exclude standard classes of X-ray emitters (late-type stars, AGN and CVs). The column density is consistent with the source being located within the Carina Nebula, roughly at a distance of $\sim$\,2\,kpc. Assuming that the X-ray emission is well described by the mean parameters given above, \candzsf\ may be more luminous and youngish than the \msev\ and thus perhaps still close to its birth place. \candzsf\ is unique in the sense that it is hotter than the seven nearby sources and may represent an evolutionary missing link between the different classes of magnetars, radio-transient and radio-quiet INSs.
The accretion scenario would require an unlikely low-velocity neutron star, combined with high accretion efficiency, to account for the X-ray luminosity of \candzsf. A systematic search for pulsations is crucial in order to establish the true nature of this X-ray source. 

\begin{acknowledgements}
The work of A.M.P. is supported by FAPESP (grant 04/04950-4), CAPES (grant BEX7812/05-7), Brazil, and the Observatory of Strasbourg (CNRS), France. R.T. aknowledges financial support from INAF-ASI under contract AAE TH-058. The work of S.P. is partially funded by INTAS through grant 6-1000014-5706; S.P. is also grateful to the Observatory of Cagliari for hospitality when part of this work was completed. The authours acknowledge the use of the ATNF Pulsar Catalogue (\texttt{http://www.atnf.csiro.au/research/pulsar/psrcat}). We thank the anonymous referee for useful comments and suggestions which helped to improve the paper.
\end{acknowledgements}

\bibliographystyle{aa}
\bibliography{ins}

\end{document}